
\magnification= \magstep 1
\global\newcount\meqno
\def\eqn#1#2{\xdef#1{(\secsym\the\meqno)}
\global\advance\meqno by1$$#2\eqno#1$$}
%
\global\newcount\refno
\def\ref#1{\xdef#1{[\the\refno]}
\global\advance\refno by1#1}
\global\refno = 1
%
\hsize = 17.5 true cm
\vsize = 24 true cm
\tolerance 10000
%
\def\s#1{{\bf#1}}
\def\tr{{\rm tr}}
\def\ln{{\rm ln}}

\baselineskip 12pt plus 1pt minus 1pt
\vskip 1.5cm
\centerline{\bf PHYSICS OF THE QUARK-GLUON PLASMA}
\vskip 1cm
\centerline{Janos Polonyi}
\vskip .5cm
\centerline{\it Laboratory of Theoretical Physics}
\centerline{\it Louis Pasteur University}
\centerline{\it 3 rue de l'Universit\'e 67087\ \ Strasbourg\ \ Cedex\ \ France}
\vskip .4cm
\centerline{\it CRN, CNRS-IN2P3}
\centerline{\it 23 rue du Loess 67200\ \ Strasbourg\ \ Cedex\ \ France}
\vskip .4cm
\centerline{\it Department of Atomic Physics}
\centerline{\it L. E\"otv\"os University}
\centerline{\it Puskin u. 5-7  \ \ 1088 \ \ Budapest \ \ Hungary}
\vskip 3cm
\centerline{{\bf ABSTRACT}}
Some features of the high temperature gluonic matter,
such as the breakdown of the fundamental group symmetry by the kinetic energy,
the screening of test quarks by some unusual gluon states and the explanation
of the absence of isolated quarks in the vacuum without the help of infinities
are presented in this talk. Special attention is paid to separate
the dynamical input inferred from the numerical results of lattice gauge theory
from the kinematics.
\medskip
\centerline{\bf 1. INTRODUCTION}
\medskip
\nobreak
\xdef\secsym{1.}\global\meqno = 1
\medskip
Contemporary physics progresses towards opposite ends. On the one
hand, one presses for the discovering the fundamental
laws in their ultimate simplicity. This is the realm of subatomic
physics where the elementary building blocks of matter are studied.
On the other hand, the description of
complex systems which are made up by simple elements and rules is sought.
Ergodicity or its absence, collective phenomena and statistical physics
label this direction. The topic of the relativistic heavy ion
collisions, RHIC, represents a
meeting ground for these two directions inasmuch as it involves the
interactions of elementary particles, quarks, and the collective
phenomena, phase transition and fragmentation. It is not sure that
we shall ultimately arrive at the understanding of the remnants of an
energetic heavy ion collision but we certainly have to use a vast area
of physics and a good deal of intuition in the analysis. The attempt at
synthesis is specially useful in our situation where the furthering
of our experimental inquiry in the microscopic domain is severely
limited by economical considerations.

The interest of the community was increased considerably in RHIC after
realizing that a "new phase of matter" might be at stakes at this
experiment. It is difficult sometime not to be lost in the rich
details of this problem but we should remember that this experiment
is worth for the concentration of our resources only if it produces
more than a quantitative change in particle production, when it
helps to discover qualitatively new laws of matter. RHIC can add
news compared to the $pp$ collisions if the collective modes become
important in modifying the medium where the elementary interactions take
place.

I do not attempt to review the theoretical aspects of the strong
interactions from the point of view of the RHIC.
Instead I shall be contented to
present some thoughts about quark confinement
what I find interesting in this subject.
The confinement of quarks represents several problems in itself because there
is no other interaction which can shield the elementary constituents of
a composite state as effectively then the confining forces between quarks.
The understanding of this mechanism would remove one of the most
troublesome obstacle of particle physics. After having identified the
confining mechanism it is equally challenging to understand its
absence when the matter is under extreme environments. It turned out
that the deconfining
phase transition is fundamentally different from other phase transitions.
A brief sketch of this latter claim is the main subject of this talk.
\bigskip
\centerline{\bf 2. CONFINEMENT MECHANISMS}
\medskip
\nobreak
\xdef\secsym{2.}\global\meqno = 1
\medskip
One can distinguish soft and hard confinement mechanisms. The former stands
for the screening of a quark \ref\gribovr\ when we try to separate it from a
hadron. This screening is achieved by the creation of a
string of mesons. This phenomenon seems similar to the supercritical state
of the QED vacuum \ref\grmurafr. Such a
screening of the isolated quarks is called soft because it involves low
energies. In order to have this mechanism operative
for hadrons where $Z=2,3$ one needs strong coupling constant. This is provided
by the anti-screening of the asymptotically free gluon exchanges.
The hard confinement mechanism is what prevents the separation
of static test quarks even in the absence of the quark Dirac sea in the
pure gluonic theory. This is believed to happen by the flux-tube formation.

It is very difficult to separate the soft and the hard confinement mechanisms
in hadronic reactions. The real flux-tube is broken by the soft confining
mechanism and is only seen indirectly in the Regge trajectories and the
heavy meson spectroscopy. In lacking of experimental guidance we have to
look for theoretical starting points.
It seems reasonable that the description of the pure gluonic vacuum
is a prerequisite for the understanding of the soft confinement mechanism.
Furthermore both the generation of the strong coupling strength and the hard
confinement mechanism seem to arise from the non-perturbative long-range
modes of the pure gluonic vacuum. Only the hard confinement
mechanism is addressed here and the quark-anti quark
vacuum polarization which is responsible for the soft mechanism
will be neglected. The qualitative arguments presented below remain
valid in their presence so long as the proper ensemble is used \ref\michr.

The best evidence for the hard confinement
mechanism comes from the numerical studies of lattice QCD in the quenched
approximation, i.e. without dynamical quarks. As far as the origin of the
string tension is concerned the dual superconductor analogies, \ref\dualscr,
and the haaron-gas description, \ref\haarr, seem to provide a consistent
framework to describe the numerical experiences.
They are based on rather different
physical scenarios: While the superconductor picture uses the dual Meissner
effect the haaron-gas description refers to localization. It remains to seen
if these mechanisms which are supported by the numerical results of lattice
QCD can be brought to a common ground.
\bigskip
\centerline{\bf 3. DECONFINEMENT PHASE TRANSITIONS}
\medskip
\nobreak
\xdef\secsym{3.}\global\meqno = 1
\medskip
Instead of pursuing the issue of the confinement mechanisms we turn to the
question of deconfinement. Our starting point will be the deconfining phase
transition because its careful study should provides us the clue of
confinement and deconfinement in the same time.

The studies of QCD at high temperature became interesting because the
smallness of the running coupling constant at high temperature
suggested that the theory becomes perturbative and reveals its elementary
constituents. But the smallness of the effective coupling strength
at the average kinetic energy is not enough to make the system perturbative.
Infrared divergences continue to plague the perturbation expansion at
arbitrary high temperature. The other expectation namely that QCD
looses the confining forces, \ref\deconfr, found out to be correct.

The appearance of new degrees of freedom can be expected from the
quantum field theory model where the hadrons are treated as pointlike
particles. Such non-gauged hadron models are non-asymtotically free and
their only renormalizable version is expected to describe non-interacting
hadrons. The non-renormalizability of the interactive effective theory
implies that the model remains useful only
up to a certain finite value of the UV cut-off. Thus new physics must come
beyond this maximal cut-off scale. The new degrees of freedom, supposedly
the quarks and gluons can be made
more explicit by reaching the maximal cut-off scale not only in the energy
of an isolated reaction but instead in a large region in space for an
extended time. One expects to trigger a phase transition in this manner.
One can pump energy into a system (i) thermally, (ii) chemically or
(iii) mechanically.
Either of these methods can be used to create the new phase.

(i) The finite temperature phase transition in hadronic matter was predicted
first by Hagedorn who noted that the level density of the observed hadronic
resonances seems to follow an exponential increase with the energy,
$n_h(E)\approx e^{E/T_H}$. Thus the simple one-particle approximation
for the hadronic partition function,
\eqn\haged{Z=\int_0^\infty dEn_h(E)e^{-{E\over T}}=
\int_0^\infty dEe^{-E({1\over T_H}-{1\over T})},}
diverges for $T>T_H$. One has to provide infinite amount of energy to
raise the temperature beyond this limiting value.

The strong coupling
expansion in hamiltonian lattice gauge theory gives similar result. Consider
U(1) gauge theory for simplicity where the eigenstates of the strong
coupling hamiltonian which includes the kinetic energy only,
\eqn\strham{H=-{1\over2}\sum_{x,i}{\delta^2\over\delta A^2_i(\s x)},}
are
\eqn\eig{\Psi_{\ell_i(\s x)}[A_i(\s x)]=
e^{ig\sum_{\s x,i}\ell_i(\s x)A_i(\s x)},}
where $\ell_i(\s x)$ is an integer valued vector field which counts the
electric
excitations along the links. Let us approximate the partition function of
a static $e^+e^-$ pair by taking into account the states where the links
are at most once excited. The excitations form a string connecting
the location of the two charges due to Gauss' law. Thus the partition
function of the gauge field in the $e^+e^-$ sector is
\eqn\epem{Z=\sum_{\Gamma}e^{-{g^2\over2}L[\Gamma]},}
where the summation is over all strings, $\Gamma$, with given end points and
$L$ stands for the length of the strings. The number of strings for large
$L$ is given by $n_s(L)\approx 5^{cL}$ where $c=O(L^0)$
since each additional link to the
string can be chosen from one of five different directions. Thus we have
\eqn\stref{Z=\sum_{L=0}^\infty e^{-L({g^2\over2T}-c\ln 5)}.}
The similarity with Hagedorn's scenario is obvious with the only difference
that it is the entropy of the gluons rather than the hadrons which
makes the partition function divergent. The strong coupling picture
allows us to see what changes as the temperature is increased beyond the
critical temperature, $T_{dec}={g^2\over2c\ln 5}$. For $T<<T_{dec}$ the
partition function is dominated by the straight string connecting the sources
and the string tension is $\sigma=g^2/2$. When $T>>T_{dec}$ then they are the
long strings wandering around in space which are important and the force
between the charges is vanishing for large separation.

The existence of a phase transition
where the string tension vanishes beyond $T_{dec}$ has been confirmed by
numerical studies of gauge models with compact gauge group.
The compactness is essential since the boundlessness of the  coordinate space
of the gauge field
makes the spectrum of the electric field operator discrete. The size
of the coordinate space, $R=O({1\over g})$, is proportional to the
inverse of the level spacing of the canonical momentum operator,
$E_i(\s x)={1\over i}{\delta\over\delta A_i(\s x)}$. Thus the gap of
the excitation
spectrum of the strong coupling hamiltonian, \strham, is proportional to
$g^2$. Larger the gap, more important is the energy in the partition
function \stref\ and the typical strings fluctuate in the closer
vicinity of the shortest line connecting the $e^+e^-$ pair.

(ii) The chemical energy pump is to squeeze hadrons into a box with high
density. Imagine that hadrons as small bags containing the quarks.
As the hadrons overlap the quarks may propagate freely between the
bags which are in contact and we recover a quark gas in a "big" bag.
This is a percolation type phase transition since the formation of a
large connected region made of the individual hadrons is needed for the
deconfinement. Unfortunately
the quark determinant in the path integral is not positive definite for
non-vansishing
baryon number and the numerical methods which are based on stochastic
processes are unable to confirm the presence of such phase transition.

(iii) Observe that the
role of the baryon density in the previous case was to control the
average hadron separation. This can be reached by
changing the pressure which leads to the mechanical way of
piling up energy. It is reasonable to expect that the hadrons in a gas start to
overlap as the pressure is increased and similar deconfining phase transition
takes place. This is equivalent with the phase transition which is supposed to
be induced by the high baryon density at RHIC. The question can be raised
in the thought experiment of colliding energetic ions on
an anti-ion beam where hadron gas with vanishing baryon density is created.
There is no problem with stochastic methods in this case.
A new density matrix should be used to describe the time averages and the
preliminary numerical results indicate the presence of the
deconfining phase transition \ref\zsoltr.
\bigskip
\centerline{\bf 4. ORDER PARAMETER OF CONFINEMENT}
\medskip
\nobreak
\xdef\secsym{4.}\global\meqno = 1
\medskip
The numerical studies of the gluon system indicate that the
expectation value of the Polyakov line,
\eqn\poly{<\tr\Omega(\s x)>=<Pe^{\int_0^\beta dtA_0(\s x,t)}>
=e^{-\beta F_q}={Z_{\bar q}\over Z_0},}
is vanishing for $T<T_{dec}$ and becomes nonzero
for $T>T_{dec}$. The first equation in \poly\ is the definition,
$P$ stands for path ordering and the gauge field
is antihermitean, $A_\mu=gA_\mu^a{\lambda^a\over2i}$. There are three
different interpretations of $<\tr\Omega(\s x)>$ what I comment here briefly.

1. {\it Quark free energy:}
The expectation value of the Polyakov line which corresponds to the insertion
of the path ordered exponential into the path integral describes the
interaction of the gluons with a static external charge. Thus this expectation
value gives the free energy, $F_q$, associated to a
static quark. $Z_{\bar q}$ and $Z_0$ is the partition function for the one
antiquark and the quarkless sectors.
The role of $A_0$ in the path integration is the projection into the
physical Hilbert space. One can write the partition function
of the vacuum sector as
\eqn\part{\eqalign{Z_0
&=\tr{\cal P}_0e^{-\beta H}\cr
&=\tr\int D[\Omega(\s x)]\hat\Omega e^{-\beta H}\cr
&=\int D[\s A(\s x)]\int D[\Omega(\s x)]<\s A^\Omega|e^{-\beta H}|\s A>,}}
where ${\cal P}_0$ is the projection operator into the gauge invariant
sector and $\hat\Omega$ is the representation of the gauge transformation
$\Omega(\s x)$ in the Hilbert space. The partition function in the
presence of a test antiquark at the location $\s x_0$ is
\eqn\partq{\eqalign{Z_{\bar q}
&=\tr{\cal P}_{\bar q}e^{-\beta H}\cr
&=\tr\int D[\Omega(\s x)]\Omega(\s x_0)\tr\hat\Omega e^{-\beta H}\cr
&=\int D[\s A(\s x)]\int D[\Omega(\s x)]<\s A^\Omega|e^{-\beta H}|\s A>
\tr\Omega(\s x_0).}}
${\cal P}_{\bar q}$ denotes the projection operator into the anti-quark
sector. The non-vanishing of the expectation value \poly\
indicates that the isolated quarks are not completely suppressed.

2. {\it Center symmetry:} The center
consists of global gauge transformations which commute with the gauge
group. It is
\eqn\cent{Z_N=\{\delta_{jk}e^{i{2\pi\over3}\ell},\ell=1,\cdots,N\},}
for the group $SU(N)$. Gauge transformations act on the gluon and the quark
fields as
\eqn\gtr{A_\mu(x)\to A^\omega_\mu(x)=
\omega(x)(\partial_\mu+A_\mu(x))\omega^\dagger(x),}
\eqn\qcn{\psi(x)\to\psi^\omega(x)=\omega(x)\psi(x).}
Global center transformations leave the gauge field invariant
because they commute with the generators but change the quark field.

It is instructive to compare these relations with those for the
rotational group in three space.
In $SU(2)$ gauge theory the center is $Z_2=\{1,-1\}$ which corresponds
to rotation in the color space by angle 0 or $2\pi$. Fermions transform
according to the spinor representation and they pick up the minus sign
under rotation by $2\pi$. The gauge field is in the adjoint, vector
representation where the vectors remain invariant under rotation by $2\pi$.

We write the partition function \part\ as
\eqn\epart{Z_0=\int D[\Omega(\s x)]e^{-{\cal H}[\Omega]},}
where ${\cal H}[\Omega]=-\ln\tr\hat\Omega e^{-\beta H}$.
One can actually generalize this construction to any matrix element,
\eqn\phsme{<\s A_f|e^{-\beta H}|\s A_i>_0=\int D[\Omega(\s x)]
<\s A_f^\Omega|e^{-\beta H}|\s A_i>
=\int D[\Omega(\s x)]e^{-{\cal H}_{\s A_f,\s A_i}[\Omega]},}
with
\eqn\melft{{\cal H}_{\s A_f,\s A_i}[\Omega]
=-\ln<\s A^\Omega_f|e^{-\beta H}|\s A_i>.}

The pure gluonic amplitudes in the vacuum sector between the gluon field
eigenstates $|\s A_i>$ and $|\s A_f>$ are center invariant in a trivial manner,
$<\s A^z_f|e^{-\beta H}|\s A_i>=<\s A_f|e^{-\beta H}|\s A_i>$ and the pure
gluonic system possesses center invariance,
${\cal H}[z\Omega]={\cal H}[\Omega]$,
${\cal H}_{\s A_f,\s A_i}[z\Omega]={\cal H}_{\s A_f,\s A_i}[\Omega]$, where
\eqn\zdef{z=e^{i{2\pi\over3}}}
is a center element. The Polyakov line is a quark observable and transforms
multiplicatively, $\Omega\to z\Omega$, hence
it is an order parameter and the non-vanishing
value of \poly\ indicates the dynamical breakdown of the center symmetry.

3. {\it Fundamental group \ref\mechr:}
It is known that the wave function can be
multi-valued when the coordinate space is multiply connected and this
gives rise a phase representation of the fundamental group
of the coordinate space. The simplest example is a particle on the circle.
The trajectories on  the circle fall into homotopy classes which can
be characterized by the winding number. When the particle moves first along
a path $\gamma_1$ with winding number $\nu_1$ after then another path,
$\gamma_2$ with winding number $\nu_2$ then it performs motion along
the path $\gamma_1\gamma_2$ which has winding number $\nu_1+\nu_2$.
The multiplication between the homotopy classes gives the fundamental group
which is $Z$ the additive group of integers for the circle.
The irreducible representations
of this group can be labeled by an angle $\theta$ which controls the
multi-valuedness of the wave functions,
$\psi(\phi+2\pi)=e^{i\theta}\psi(\phi)$.
Since the observable quantities correspond to operators which have
single valued matrix elements the linear space, ${\cal H}_\theta$,
consisting of the wave functions with a given
$\theta$-parameter are closed with respect to the action of the physical
operators and form a superselection class.
The phase $\theta$ is the magnetic flux going through the circle and its
appearance in observable quantities is the Aharonov-Bohm effect. One should
bear in mind that the hamiltonian and some other observable are actually
$\theta$ dependent.

Another example is provided by rotations in three space. Consider the wave
function of an N-body system expressed by the help of the collective
coordinates,
\eqn\mbwf{\psi(\s x_1,\cdots,\s x_N)=e^{-i\s X\s P}
{\cal D}^j_{m_1,m_2}(\phi,\theta,\chi)\eta(r_{rel}),}
where $\s X$ is the center of mass coordinate $\phi,\theta,\chi$ are the
Euler angles connecting the laboratory and a body fixed coordinate system
and $r_{rel}$ stands for the set of relative coordinates.
There is nothing inconsistent in specifying half integer values for $j$ in the
irreducible rotation matrix elements ${\cal D}^j_{m_1,m_2}$ since the
rotational group $SO(3)=SU(2)/Z_2$ is doubly connected.

This is the point which
generalizes for $SU(N)$: The gauge field transform according to the adjoint
representation of the global gauge group, $\s A\to\omega\s A\omega^\dagger$,
c.f. \gtr. The center elements commute with the generators and are
represented trivially on the adjoint vectors. The space of global
gauge transformation is thus $SU(N)/Z_N$. This is a multiply
connected space with the fundamental group $Z_N$. Thus the wave function
which depends on the gauge field may be N-valued. The label of this phase
representation of the center is the N-ality, which is the triality for QCD.
Summarising: the center of the global gauge group is the fundamental group
of the global gauge symmetry.

The formal global symmetries of the system may be broken dynamically
under certain circumstances. When this happens the charge with respect to
the symmetry group ceases to be a good quantum number and the physical
states become linear superpositions of components with different charges.
The charge of the fundamental group is the multi-valuedness angle, N-ality.
When the fundamental group symmetry is broken dynamically then the
wave functions of the physical states change not only their phase but
the absolute magnitude as well under fundamental group transformation.

The last remark about this point is an extension of a technical point.
Consider the many-body system \mbwf\ in a simple connected space.
One can always, as was noted, use half-integer $j$ in constructing a
wave function. Since $j(j+1)$ is the eigenvalue of the total angular momentum
square, $\s J^2$, and the total angular momentum is the sum of the angular
momenta of the particles, $\s J=\sum_a\s L_a$, at least one particle must
carry half integer orbital angular momentum when $j$ is half-integer.
Thus there must be a way to describe half-integer
angular momentum states with double-valued wave function in simply connected
space. The solution is straightforward and simple. The
sufficient condition is that the interaction excludes the particles from
certain region in such a manner that the left over of the phase space
is multiply connected. Such an interaction can generate multi-valued
wave functions even in simply connected space.

After these preliminary remarks we are
in the position to give the third interpretation of \poly.
The density matrix for gluons in the presence of an isolated heavy anti-quark
is
\eqn\densga{\rho_{\bar q}[\s A,\s A']
=<\s A|{\cal P}_{\bar q}e^{-\beta H}|\s A'>=\int D[\Omega(\s x)]
<\s A^\Omega|e^{-\beta H}|\s A'>\tr\Omega(\s x_0)}
in the field diagonal representation. The factor $\tr\Omega(\s x_0)$ is
to guarantee Gauss' law at the location of the test charge. By the
help of the invariance property of the group integration,
$\int dgf(g)=\int dgf(gh)$, one can easily prove the relation
\eqn\mvdm{\rho_{\bar q}[\s A^z,\s A']
=e^{-i{2\pi\over3}}\rho_{\bar q}[\s A,\s A'].}
This shows that the gluon wave functional which corresponds
to an isolated quark is multi-valued. The non-vanishing of \poly\
is just the statement that the gluon states with multi-valued wave functional
contribute in the high temperature phase. Had the integration in \part\ or
\partq\ had been performed completely \poly\ would have been vanishing
and the states with multi-valued wave functions would have been absent.
The appearance of such states in the partition function indicates that the
gluons are constrained into a multiply connected region of the
configuration space in the high temperature phase and the fundamental group
symmetry is broken dynamically.
\bigskip
\centerline{\bf 5. RIDDLES}
\medskip
\nobreak
\xdef\secsym{5.}\global\meqno = 1
\medskip
The following arguments are to demonstrate some of the unusual features
of the deconfining phase transition:

1. {\it High energy symmetry breaking:} The only way to violate a
formal symmetry by the dynamics is the spontaneous symmetry breaking.
Then the symmetry is broken in the vacuum, at low energy and gets
restored at high energy. The center symmetry is broken
at high temperature and is restored at low temperature. What is the
mechanism of this unusual symmetry breaking pattern ?

2. {\it Asymptotic states:}
One may compare the deconfining phase transition with the insulator-conductor
Mott transition in solid state physics. In the latter the lattice impurities
make the electron wave function localized at high temperature and lead
to the breakdown of the conductance.
Note that the basic difference between localization and confinement
is that while the former allows charged states with zero momentum the latter
does not. The thermal averages,
\eqn\qsma{<A>_{th}=\sum_np(E_n)<n|A|n>,}
are computed in Quantum Statistical Mechanics by the help of the eigenstates
of the hamiltonian, $H|n>=E_n|n>$, and the distribution function, $p(E)$.
There are extended electron states on the solid state lattice which provide
the main contribution to the partition function in the insulator phase. But
there are no finite energy eigenstates of the QCD hamiltonian with net
color charge. Where do the deconfined quarks of the high temperature phase
come from in QCD\nobreak?

3. {\it Finite volume effects:} According to the second interpretation of the
Polyakov line it is the order parameter of a formal symmetry of the partition
function and such must be preserved for a finite system. Thus
\poly\ is vanishing when the system is placed into a finite volume
with open boundary condition in the spatial directions. The
first interpretation then assigns infinite energy for the static test
charge. Since the phase transition does not changes the ultraviolet
properties of the system the divergence in the energy can not be
ultraviolet. The finite quantization box provides an infrared cut-off.
How can then the energy of a static charge be infinite in a finite box ?
\bigskip
\centerline{\bf 6. SOLUTIONS}
\medskip
\nobreak
\xdef\secsym{6.}\global\meqno = 1
\medskip
1. {\it High energy symmetry breaking:} Though we started with the partition
function at finite temperature the situation is formally similar when a
general matrix element, \phsme, is considered. Actually,
both the partition function and the matrix element show the
phase transition for short time in the mean field approximation. To see this
first note that the potential energy is negligible compared to the kinetic
term for the short time processes. The kinetic energy of the
temporal gauge hamiltonian is quadratic in the gauge field and
is large if the initial and the final configurations differ for short
time. By comparing \melft\ with
\eqn\qmme{-\ln<\s x_f|e^{{\beta\over2}{\partial^2\over\partial\s x^2}}
|\s x_i>={1\over2}\ln(2\pi\beta)+{(\s x_f-\s x_i)^2\over2\beta},}
one sees that ${\cal H}[\Omega]$ is minimal for $\Omega(\s x)=z^k$
because this is the case when the initial and the final configurations
agree and the static configuration with vanishing kinetic energy
dominates the path integral. When the
initial and the final configurations differ in the short time limit then
the large kinetic energy makes the transition suppressed and
${\cal H}[\Omega]$ increases. The maximum of ${\cal H}[\Omega]$
is reached in the middle between two center elements,
$\Omega=\sqrt{z}$. The free system has no internal scale
so the energy barrier between the minima is proportional to $V\beta^3$
where $V$ is the three volume. The system is trapped in one of the
center minima for short time. Similar periodic structure is found for
${\cal H}_{\s A_f,\s A_i}[\Omega]$ with minima at
$\Omega(\s x)=z^k\tilde\Omega(\s x)$
where $\tilde\Omega(\s x)$ minimizes
$-\tr\int d^3x(\s A^\Omega_f(\s x)-\s A_i(\s x))^2.$ This simple argument
shows that the fundamental symmetry is broken by the kinetic energy
for the amplitudes of short time processes. This symmetry breaking mechanism
is fundamentally different than the spontaneous symmetry breaking which
is driven by the potential energy and is effective at low energy,
long time. The breakdown of the fundamental group symmetry and
the applicability of the mean field solution provides the
basis of the parton picture in high energy hadron physics.

2. {\it Asymptotic states:} The last equation in \poly\ expresses
the fact that the partition function of the sector which comprises
that states with a color charge of an anti-quark is non-vanishing.
What states contribute to this partition function ? Since the system
in question consists of gluons these states belong to the gluonic
Fock space. They share an unusual property namely that the total
charge seen by local operators in these state is that of an anti-quark.
The apparent contradiction that the sum of integers (gluon charge)
can not be $1\over3$ (quark charge) is circumvented by piling up
infinitely many gluons. The limit of an infinite
series can be modified by reshuffling the series. An example
of a state which has infinitely many plane wave components and
belong to the anti-quark charge sector is
\eqn\exampl{|\s A(\s x)>_{\bar q}=\int d\Omega_0
(\Omega_0)_{ab}|\s A^{\Omega_0}(\s x)>,}
where $\s A(\s x)$ is an arbitrary localized non-vanishing gluon field
configuration, $d\Omega_0$ is the invariant integration over global
gauge transformations and $(\Omega_0)_{ab}$ is a matrix element of
the integral variable. Note that by replacing $(\Omega_0)_{ab}$
by any other matrix element of an irreducible representation of the
gauge group the resulting nontrivial state belongs to the given
representation. This is the demonstration of the well known fact
in Quantum Mechanics that the coordinate eigenstate has non-vanishing
projection in all angular momentum subspace. In a similar manner,
the gluon state $|\s A(\s x)>$ contains components from each charge
sector.

It is easy to see that the wave functional of
\exampl\ is N-valued. Thus the gluonic state surrounding the test quark
carries the color charge of an anti-quark and the gauge invariance of \poly\
assures that the dressed test quark is color neutral.
The following picture emerges for the deconfining phase transition:
At low temperature where the $\Omega(\s x)$, i.e. end point dependence
is weak in the right hand side of \phsme\ the $\Omega(\s x)$
integration is properly performed. This reinforces Gauss' Law and
makes the gluonic states neutral in the vacuum sector. At high temperature
the infinite dimensional $\Omega(\s x)$ integration is not carried
out properly. The strong end point dependence restricts the integration
into a smaller region. This is formally similar to the spontaneous breakdown
of symmetries except that it refers now a given matrix element, \phsme,
instead of the vacuum. When the projection into the color singlet
sector is not implemented properly then states with ill-defined charge
show up in the partition function. Since it is the center of the gauge
group which breaks down dynamically only the fractional part of
the color charge becomes ill-defined. The expectation value \poly\
picks up the contribution of the anti-quark charge components of the
gluon states. Thus the deconfining phase transition is the point where
a constraint what we impose formally is rejected by the dynamics.

One is attempted to speculate about the following toy-model where a quark
flavor, say the top quark, is missing from the electro-weak current.
A deep inelastic lepton-hadron collision made on a hadron with non-vanishing
top quantum number is interpreted in the framework of this model as
the finding of a deconfined quark. Indeed, the top quark which screens the
other quark(s) of the hadron is not seen in the experiment. Thus it would have
been more appropriate to use the word "screened" rather than "deconfined" for
the
high temperature phase of the pure gluon system. The quarks can be isolated
in this phase only because there are unusual gluon states available
which can provide complete screening.

It is worth while noting that the hamilton operator is given as
\eqn\hamaq{H_{\bar q}=H{\cal P}_{\bar q}=\int D[\Omega(\s x)]
H\hat\Omega\tr\Omega(\s x_0)}
in the anti-quark sector. This expression is obtained from \densga\ by
taking into account $[H,{\cal P}_{\bar q}]=0$. The complex phase,
$\tr\Omega(\s x_0)$, has an effect which is similiar to that of the
Wess-Zumino-Witten term in the Skyrme model \ref\wzwr. It attaches a phase
factor to global internal space rotations in such a manner that the
hedgehog configurations, the chromomagnetic monopoles in our case,
acquire fermi statistics \mechr. The gluonic state which screens
the deconfined quark can be written as a linear superposition,
$\Psi[\s A(\s x)]=\Psi_0[\s A(\s x)]+\Psi_1[\s A(\s x)]$, where
the support of $\Psi_n[\s A(\s x)]$ is in the space with even or
odd number of monopoles for $n=0$ and $n=1$, respectively. Thus the
deconfined quark is such a composite particle of the quark and the screening
gluonic cloud that these two components obey
boose ($n=1$) and fermi ($n=0$) statistics.

An important question hiding in this scenario is whether the unusual
gluon states which can screen a quark are stable. The screening cloud
around a charge in the electron gas is created by the original charge
and disappears in its absence. The strong non-linearity of the gluonic
system in particular the instanton solution may make the screening states
stable at least semiclassically. If the screening states prove to be
stable then the phenomenology of the high temperature gluonic phase should
be guided by a simple quark model for these states.

{\it Finite volume effects:}  This puzzle will be resolved in two
steps. The fundamental idea is to explain the vanishing of a quantity like
$ce^{-\beta F}$ by the vanishing of the coefficient $c$ rather than using
the rather problematical $F=\infty$. We start with the well known
two slits thought experiment in Quantum Mechanics. Place a particle
source and a detector on the two opposite sides of a wall with two small
slits. The location of the detector is chosen in such a manner that
the difference of the two paths along which the particles can propagate
is a half wave length. The amplitude of finding a particle at the
detector is $\psi=c_1e^{-itE_1/\hbar}+c_2e^{-itE_2/\hbar}$. It is vanishing
because $c_1=-c_2$ and $E_1=E_2$. How could we give account of the
absence of the particles at the location of the detector without
knowing about the interference between the allowed paths\nobreak?
The amplitude of finding a particle can generally be written as
$\psi=ce^{-itE/\hbar}$ where $\hbar$ has an infinitesimally small
imaginary part to comply causality. In lacking any specific kinematical reason
to set $c=0$ we end up arguing for $E=\infty$. But the
information that the amplitude is vanishing due to a destructive
interference makes it possible to explain the measurement without
invoking infinities.

This scenario can be carried over to the internal space of the gluons.
The vanishing of \poly, more precisely \partq, can be understood in the
following manner: Imagine the path integral representation for the
matrix element $<\s A^\Omega|e^{-\beta H}|\s A>$ in the right hand side.
We introduce equivalence
classes for the configurations $\s A(\s x,t)$ of the path integral. The
configurations $\s A(\s x,t)$ and $\s A'(\s x,t)$ belong to the same class
if their end points can be made equal by a time dependent global gauge
transformation, i.e. when
\eqn\equiv{\cases{\eqalign{\s A^\omega(\s x,0)&=\s A'(\s x,0)\cr
\s A^\omega(\s x,\beta)&=\s A'(\s x,\beta)}}}
hold for an appropriate
$\omega(\s x,t)=\omega(t)$. Each equivalence class can further be
divided into $N$ components. This is because $\omega(0)$ and
$\omega(\beta)$ must be center elements in order to satisfy \equiv\
for $\s A=\s A'$ and $\omega(t)$ and $\omega(t)\omega^\dagger(0)$ have the same
effect on the gauge field. There is a destructive interference
in \partq\ between the components within each equivalence class.

To prove this first note that the gluonic matrix element,
$<\s A^\Omega|e^{-\beta H}|\s A>$, depends on the end points of the
trajectories. Since the global center transformation does not change
the gauge field this amplitude
and the integral measure is the same in each
component. The Polyakov line in \partq, $\Omega(\s x_0)$, which takes into
account the propagation of the test quark gets multiplied by a center element
as we move from one component to another. So the sum of the
components is proportional to
\eqn\cancell{\sum_{j=1}^Ne^{i{2\pi\over N}j}=0.}
The detector which is prepared to identify an isolated quark remains
silent because whenever a quark arrives there it comes through $N$
different paths with destructing interference.
The large kinetic energy suppresses $N-1$ paths in the amplitude for
short processes and the quarks can be isolated.
\bigskip
\centerline{\bf 7. DYNAMICS V.S. KINEMATICS}
\medskip
\nobreak
\xdef\secsym{7.}\global\meqno = 1
\medskip
It is important to distinguish the kinematical constructions from the
dynamical assumptions. The only dynamical input to support the arguments
above is the non-vanishing of \poly\ for $T>T_{dec}$.
The non-vanishing of an order parameter is always the result of the
reduction of the region of an integration, the system being trapped in a
part of the whole phase space. How does this happen here ?
As mentioned in the previous Section the integration in \poly\ is
restricted in our case. The formal functional integral is over the space
${\cal G}=\prod_{\s x}G(\s x)$
which is the direct product of the gauge group, $G(\s x)$, at each lattice
site.
The local symmetries are never broken dynamically so one has to separate off
the
global gauge group, $G_0$, and its center, $C$, by writing
\eqn\symmgr{{\cal G}=C\otimes G_0/C\otimes{\cal G}/G_0.}
The integration is not carried out for the first factor in \poly\
above the critical temperature according to the numerical simulations.
The second factor is an N-fold connected space which is the source of the
N-valuedness of the gluonic wave functional.

A functional integral which is of infinite
dimensional can in principle undergo phase transitions. The
restricted "symmetry broken" realization of the integral corresponds
to a stable dynamical system if the domain of integration is closed
under the infinitesimal deformations of the trajectories. This is
because the equation of motion and other constraints, such as the Ward
identities which express the presence of formal continuous symmetries
are written by the help of the canonical coordinate and momentum operator,
$\s A(\s x)$, and $\s E(\s x)$, respectively which are multiplicative
and derivative operators. The former is diagonal in the coordinate
eigenstate basis and the latter generates infinitesimal displacements
in the coordinate eigenvalues. Thus the insertion of these operators
do not lead out of the domain of integration and consequently can be
taken into
account properly if the space of trajectories consists of complete homotopy
classes.

On the one hand, the configuration space of the gluon system, i.e. the space
of field configurations $\s A(\s x)$ is obviously simply connected. But on
the other hand, the dynamics may exclude
the system from certain region of the configuration space in such a manner
that the left over is multiply connected. In our case this amounts
to the condition that the internally paralel field configurations,
$\s A^{(p)}(\s x)=\s A\alpha(\s x)$,
are suppressed by the dynamics in the high temperature phase,
\eqn\dync{\rho[\s A^{(p)},\s A^{(p)}]=0.}
In fact, \dync\ is needed to make the $N$ components introduced after eq.
\equiv\ disconnected. The suppression of the long range field is
the result of the non-perturbative magnetic screening in the high temperature
phase.
\bigskip
\centerline{\bf 8. CONCLUSIONS}
\medskip
\nobreak
\xdef\secsym{8.}\global\meqno = 1
\medskip
Algebraic and topological arguments were presented to demonstrate some
unusual features of the high temperature phase of the gluon system.
In particular, a new type of dynamical symmetry breaking which is driven by
the kinetic energy, a screening mechanism by unusual gluonic states and
an explanation of the absence of isolated quarks in the vacuum without
evoking infinite energy was given. Though the arguments were exact,
i.e. do not involve approximations they remained qualitative only.
The detailed quantitative verification of these points with special
attention paid to the dynamical quarks is needed in order to fit
the phenomena mentioned above into the usual description of the
quark-gluon plasma which is based on the partially resumed perturbation
expansion.
\bigskip
\centerline{\bf REFERENCES}
\medskip
\nobreak
\def\pr{{\it Phys. Rev.\ }}
\def\pre{{\it Phys. Rep.\ }}

\def\np{{\it Nucl. Phys.\ }}
\def\pl{{\it Phys. Lett.\ }}

\def\gribov{V. N. Gribov, J. Nyiri, {\it Supercritical Charge in
Bosonic Vacuum} LU-TP-91-15, unpublished.}
\def\grmuraf{W. Greiner, B. M\"uller, J. Rafelski, {\it Quantum
Electrodynamics of Strong Fields}, Springer-Verlag, Berlin, 1981.}
\def\dualsc{S. Mandelstam, \pre {\bf 23C} 245 1976;
G. t' Hooft, \np {\bf B190} 455 1981;
H. B. Nielsen, P. Olesen, \np {\bf B61} 45 1976.}
\def\mich{M. Oleszczuk, J. Polonyi, {\it Canonical versus Grand Canonical
Ensembles in QCD}, submitted to {\it Z. Phys.};
M. Faber, O. Borisenko, G, Zinovjev, {\it Triality in QCD at Zero
and Finite Temperature: a New Direction}, to appear in \np.}
\def\haar{K. Johnson, L. Lellouch, J. Polonyi, \np {\bf B367} 675, 1991.}
\def\deconf{L. Susskind, \pr {\bf D20} 2610 1979; A. M. Polyakov,
\pl {\bf 72B} 447 1978.}
\def\zsolt{J. Polonyi, Zs. Schram, {\it Nucl. Phys. Proc. Suppl.} {\bf B42}
926 1995.}
\def\mech{J. Polonyi, in {\it Quark-Gluon Plasma}, edt. R. Hwa,
World Scientific, 1988.}
\def\wzw{E. Witten, \np {\bf B223} 422 1983, ibid 433 1983.}
\item{\gribovr}\gribov
\item{\grmurafr}\grmuraf
\item{\michr}\mich
\item{\dualscr}\dualsc
\item{\haarr}\haar
\item{\deconfr}\deconf
\item{\zsoltr}\zsolt
\item{\mechr}\mech
\item{\wzwr}\wzw
\vfill
\eject
\end